\documentclass[12pt,preprint]{aastex}
\usepackage{psfig}

\def\degr{\hbox{$^\circ$}}
\newcommand\Lunit   {ergs s$^{-1}$}

\begin{document}

\title{
{\itshape Chandra} observations of the accretion-driven millisecond
X-ray pulsars XTE J0929--314 and XTE J1751--305 in quiescence}

\author{
Rudy Wijnands\altaffilmark{1}, Jeroen Homan\altaffilmark{2}, Craig
O. Heinke\altaffilmark{3,4,5} Jon M. Miller\altaffilmark{3,6}, Walter
H. G. Lewin\altaffilmark{2}}

\altaffiltext{1}{Astronomical Institute ``Anton Pannekoek'',
University of Amsterdam, Kruislaan 403, 1098 SJ, Amsterdam, the
Netherlands; rudy@science.uva.nl}

\altaffiltext{2}{Center for Space Research, Massachusetts Institute of
Technology, 77 Massachusetts Avenue, Cambridge, MA 02139, USA;
jeroen@space.mit.edu, lewin@space.mit.edu}

\altaffiltext{3}{Harvard-Smithsonian Center for Astrophysics, 
60 Garden Street, Cambridge, MA 02139, USA;
cheinke@head.cfa.harvard.edu; jmmiller@head.cfa.harvard.edu}

\altaffiltext{4}{Northwestern University, Dept. of Physics \&
  Astronomy, 2145 Sheridan Rd., Evanston, IL 60208}

\altaffiltext{5}{Lindheimer Postdoctoral Fellow}

\altaffiltext{6}{NSF Astronomy \& Astrophysics Fellow}

\begin{abstract}

We observed the accretion-driven millisecond X-ray pulsars XTE
J0929--314 and XTE J1751--305 in their quiescent states using {\it
Chandra}.  From XTE J0929--314 we detected 22 source photons (in the
energy range 0.3--8 keV) in $\sim$24.4 ksec, resulting in a
background-corrected time-averaged count rate of $9\pm2 \times
10^{-4}$ counts s$^{-1}$.  The small number of photons detected did
not allow for a detailed spectral analysis of the quiescent spectrum,
but we can demonstrate that the spectrum is harder than simple thermal
emission which is what is usually presumed to arise from a cooling
neutron star that has been heated during the outbursts. Assuming a
power-law model for the time-averaged (averaged over the whole
observation) X-ray spectrum, we obtain a power-law index of
$1.8^{+0.6}_{-0.5}$ and an unabsorbed X-ray flux of $6^{+4}_{-2}
\times 10^{-15}$ ergs s$^{-1}$ cm$^{-2}$ (for the energy range 0.5--10
keV), resulting in a 0.5--10 keV X-ray luminosity of $7^{+5}_{-2}
\times 10^{31}$ ($d$/10 kpc)$^2$ ergs s$^{-1}$, with $d$ the distance
toward the source in kpc.  No thermal component could be detected;
such a component contributed at most 30\% to the 0.5--10 keV flux.
Variability in the count rate of XTE J0929--314 was observed at the
95\% confidence level. We did not conclusively detect XTE J1751--305
in our $\sim$43 ksec observation, with 0.5--10 keV flux upper limits
between 0.2 and 2.7 $\times 10^{-14}$ ergs s$^{-1}$ cm$^{-2}$
depending on assumed spectral shape, resulting in 0.5--10 keV
luminosity upper limits of 0.2 -- 2 $\times 10^{32}$ ($d$/8 kpc)$^2$
ergs s$^{-1}$.  We compare our results with those obtained for other
neutron-star X-ray transients in their quiescent state, and in
particular with the quiescent properties of SAX J1808.4--3658. Using
simple accretion disk physics in combination with our measured
quiescent luminosity of XTE J0929--314 and the luminosity upper limits
of XTE J1751--305, and the known spin frequency of the neutron stars,
we could constrain the magnetic field of the neutron stars in XTE
J0929--314 and XTE J1751--305 to be less than $3\times 10^9 {d \over
{\rm 10~kpc}}$ and $3 - 7 \times 10^8 {d
\over {\rm 8~kpc}}$ Gauss (depending on assumed spectral shape of the
quiescent spectrum), respectively.

\end{abstract}


\section{Introduction}

Neutron stars in low-mass X-ray binaries accrete matter from solar
mass companion stars. The sub-group of neutron-star transients spend
most of their time in quiescence during which hardly any or no
accretion occurs onto their neutron stars. However, these transients
sporadically become very luminous in X-rays ($>$$10^{36-38}$~\Lunit)
owing to a huge increase in the accretion rate. During those
outbursts, these sources can be readily studied with the available
X-ray instruments, but so far only about a dozen sources have been
studied in their much dimmer quiescent states. In quiescence, they
typically exhibit 0.5--10 keV luminosities of $10^{32-34}$~\Lunit~and
their spectra are usually dominated by a soft component which can be
described by a thermal model (either a black-body or a neutron-star
atmosphere model). This emission is generally ascribed to the cooling
of the neutron star which has been heated during the outbursts (e.g.,
via deep crustal heating; Brown, Bildsten, \& Rutledge 1998). For
several quiescent systems an additional power-law shaped component is
present in their X-ray spectra and it dominates above a few keV (e.g.,
Asai et al. 1998; Rutledge et al. 2001). The origin of this power-law
component is not understood but it has been proposed to be due to
residual accretion on the neutron-star magnetic field, an active
pulsar mechanism, or shock emission due to the interaction of the
pulsar wind and matter which is still being transferred from the
companion star (e.g., Stella et al.~1994; Campana et al.~1998; Campana
\& Stella 2000).

The fractional contribution of this power-law component to the flux in
the 0.5--10 keV energy range varies significantly between systems. In
some systems this component cannot be detected (with less than 10\% of
the 0.5--10 keV flux possibly due to such a component), but in other
systems it contributes up to half the emission in the 0.5--10 keV
energy range (e.g., Asai et al.~1998; Rutledge et
al.~2001). Currently, only two quiescent systems (SAX J1808.4--3658
and EXO 1745--248) have been found to exhibit quiescent spectra which
are fully dominated by the power-law component (with over 90\% of the
0.5--10 keV flux due to the power-law component; Campana et al.~2002;
Wijnands et al.~2004). For several other systems (i.e., SAX
J1810.8--2609 and XTE J2123--058; Jonker, Wijnands, \& van der Klis
2004a; Tomsick et al.~2004) it has also been found that the quiescent
spectra could be adequately fitted with only a power-law
model. However, the statistics of the data for those sources were
rather limited, still allowing for a thermal component which
contributed more than 50\%--70\% to the 0.5--10 keV flux.

During outburst SAX J1808.4--3658 is an accretion-driven millisecond
X-ray pulsar and its anomalous X-ray properties in quiescence (its
hard quiescent spectrum and low luminosity of $\sim$$5\times 10^{31}$
ergs s$^{-1}$; Campana et al.~2002) might be related to the expected
higher magnetic field strength of the neutron star in this system,
compared to that of the neutron stars in the non-pulsating
systems. However, the recent study of the neutron-star X-ray transient
EXO 1745--248 located in the globular cluster Terzan 5 during its
quiescent state has cast doubt on this hypothesis. Wijnands et
al.~(2004) found that the Terzan 5 system, which is not observed to
pulsate during outburst, also had a quiescent spectrum which was fully
dominated by the power-law component (more than 90\% of the 0.5--10
keV flux was due to the power-law component) just like SAX
J1808.4--3658. However, EXO 1745--248 had a quiescent luminosity
($\sim$$2\times10^{33}$ ergs s$^{-1}$) which was a factor of $\sim$40
higher than what has been observed for SAX J1808.4--3658.

The physical process(es) behind the power-law component and the
differences among sources in terms of the properties of this spectral
component are unknown.  However, some clues about the nature of this
spectral component were recently glimpsed in the work by Jonker et
al.~(2004a, b). They compared all quiescent neutron star X-ray
transients for which spectral information was available and found that
the fractional contribution of the power-law component to the 0.5--10
keV fluxes is lowest when these sources have quiescent luminosities of
$\sim 1 - 2 \times 10^{33}$ ergs s$^{-1}$. They found that both at
higher and lower quiescent luminosities the fractional contribution of
the power-law component to the 0.5--10 keV flux increases. So far,
only EXO 1745--248 did not follow this correlation. It remains to be
seen whether this transient is an unusual system or if the correlation
found by Jonker et al.~(2004a, b) is spurious and more sources similar
to EXO 1745--248 will be found.

In recent years, four additional accreting millisecond X-ray pulsars
(after SAX J1808.4--3658) have been discovered (Markwardt et al.~2002;
Galloway et al. 2002; Markwardt, Smith, \& Swank 2003; Markwardt \&
Swank 2003).  Those systems are prime targets to observe in quiescence
to test the hypothesis that the unusual quiescent properties of SAX
J1808.4--3658 are due to a relatively strong neutron-star magnetic
field strength compared to that of the non-pulsating systems: one
would expect that the magnetic fields of the neutron stars in those
additional systems are of similar strength as that of the neutron star
in SAX J1808.4--3658. To this end we had observations of XTE
J1751--305 and XTE J0929--314 scheduled during cycle 5 of {\it
Chandra}. Here we report on the results of those observations.

The second accretion driven millisecond X-ray pulsar, XTE J1751--305,
was discovered on April 3, 2002 (Markwardt et al.~2002). It was found
to harbor a neutron star with a spin frequency of 435 Hz which is
located in a binary system with an orbital period of 42 minutes
(Markwardt et al.~2002).  The third accreting millisecond X-ray
pulsar, XTE J0929--314, was discovered by Remillard (2002) on April
13--18, 2002, using the all-sky monitor (ASM) aboard the {\it Rossi
X-ray Timing Explorer} ({\it RXTE}). Only later (May 2, 2002), when
the proportional counter array aboard {\it RXTE} observed the source,
was it discovered to be an accreting millisecond X-ray pulsar with a
pulse frequency of 185 Hz (Remillard, Swank, \& Strohmayer 2002). The
orbital period of the system was found to be 44 minutes (Galloway et
al.~2002).

\section{Observations, analysis, and results}

We observed XTE J0929--314 with {\it Chandra} on March 18, 2004,
between 02:30 and 10:02 UT (resulting in an exposure time of
$\sim$24.4 ksec) and XTE J1751--305 on June 26, 2004 between 07:59 and
21:12 UT (resulting in an exposure time of $\sim$43.0 ksec) We used
the ACIS-S3 CCD and we selected a 1/4 sub-array to limit the pile-up
in case the source fluxes exceeded $\sim$10$^{-12}$ ergs s$^{-1}$
cm$^{-2}$ (as we will show below, both sources had quiescent fluxes
significantly lower than this flux level so no pile-up occurred during
our observations). We checked for background flares during our
observations, but none were found allowing us to use all available
data. Our analyzes were performed using the CIAO software package
(version 3.1) and the standard {\it Chandra} analysis
threads\footnote{See http://cxc.harvard.edu/ciao/ for CIAO and the
analysis threads}.
 
\subsection{XTE J0929--314}

\subsubsection{Image analysis}

\begin{figure}
\begin{center}
\begin{tabular}{c}
\psfig{figure=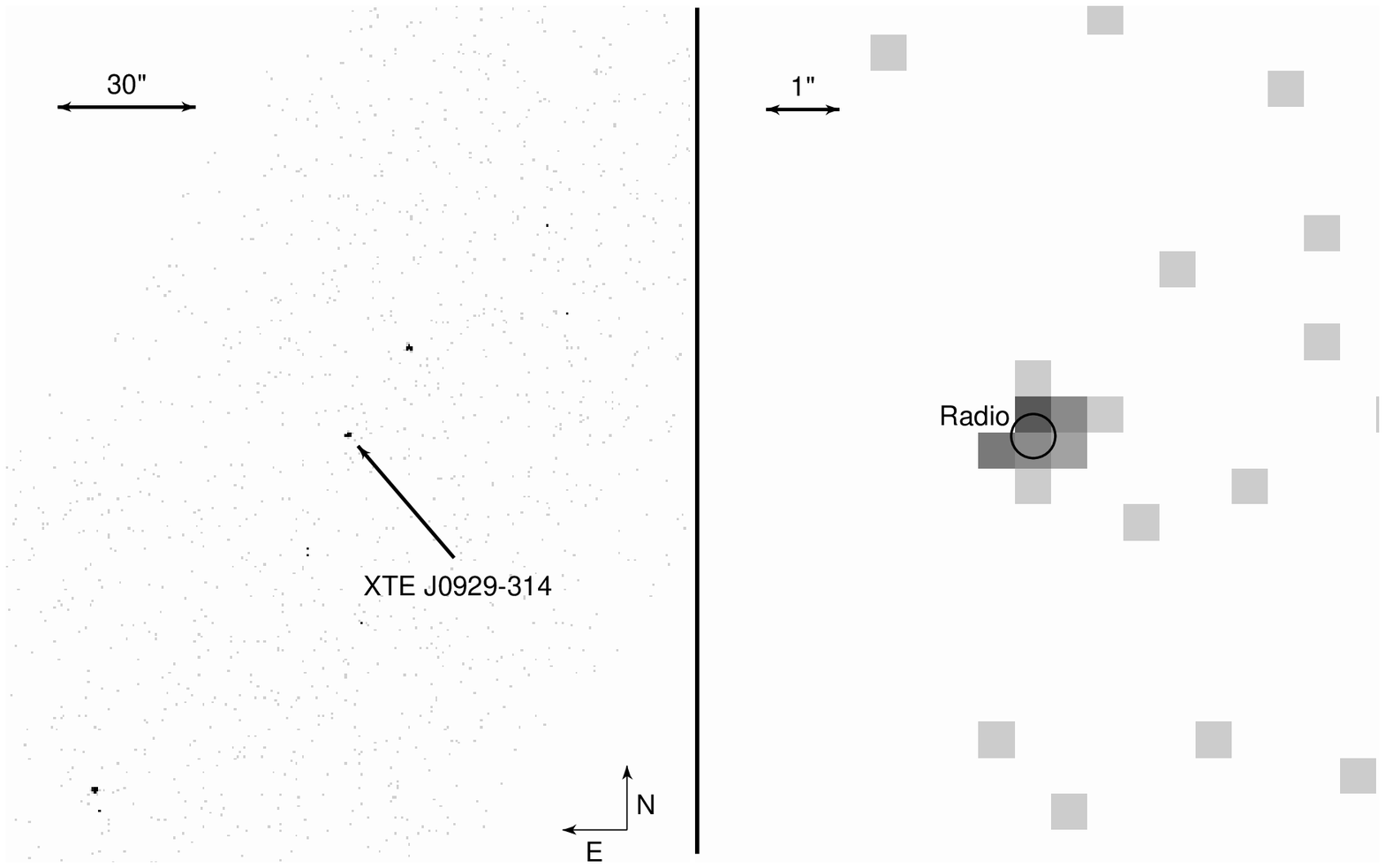,width=16cm}
\end{tabular}
\figcaption{
The {\it Chandra} image of the field around XTE J0929--314 (left) and
a close-up of XTE J0929--314 (right) with the radio error circle for
the source obtained during outburst (Rupen et al.~2002). In both
figures, north is up and east is to the left. In the left panel the
arrow indicates which source is XTE J0929--314. The source located to
the north-west of XTE J0929--314 is CXOU J092919.1--312244 and the
source to the south-east (in the bottom left corner) is CXOU
J092924.4--312420.
\label{fig:image} }
\end{center}
\end{figure}

We used the tool 'wavdetect' to search for point sources in our data
of XTE J0929--314 and to obtain the coordinates of each source which
was detected. For XTE J0929--314, we used the tool 'dmextract' to
extract the number of observed photons and the count rate (for the
energy range 0.3--8 keV). As source extraction region we used a circle
centered on the source position (as obtained with 'wavdetect') with a
radius of 1.5$''$. For background extraction region we used an annulus
centered on the source position with an inner radius of 5$''$ and an
outer radius of $20''$ (a larger outer radius could not be used since
the field source CXOU J092919.1--312244 was within $25''$ of XTE
J0929--314 as can be seen in Fig.~\ref{fig:image} left panel). We
detected four sources in our image, including XTE J0929--314; we will
not discuss the other three sources further in this article. Instead
we concentrate on the observed X-ray properties of XTE J0929--314 in
its quiescent state. The coordinates we obtained for XTE J0929--314
are right ascension $09^h ~29^m ~20^s.180$ and declination $-31\degr
~23' ~03''.5$ (epoch J2000.0), which are consistent with the
coordinates of the source obtained during outbursts in the X-ray
(Juett, Galloway, \& Chakrabarty 2003), optical (Greenhill, Giles, \&
Hill 2002), and radio bands (Rupen, Dhawan, \& Mioduszewski 2002; see
Fig.~\ref{fig:image} right panel).  The total number of source photons
detected in the energy range 0.3--8 keV from XTE J0929--314 is 22,
resulting in a net time-averaged count rate of $9\pm2 \times 10^{-4}$
counts s$^{-1}$ (after background subtraction; less than 0.4
background photons are expected in our source extraction region).

\subsubsection{Count rate analysis \label{subsubsection:cr_0929}}

\begin{figure}
\begin{center}
\begin{tabular}{c}
\psfig{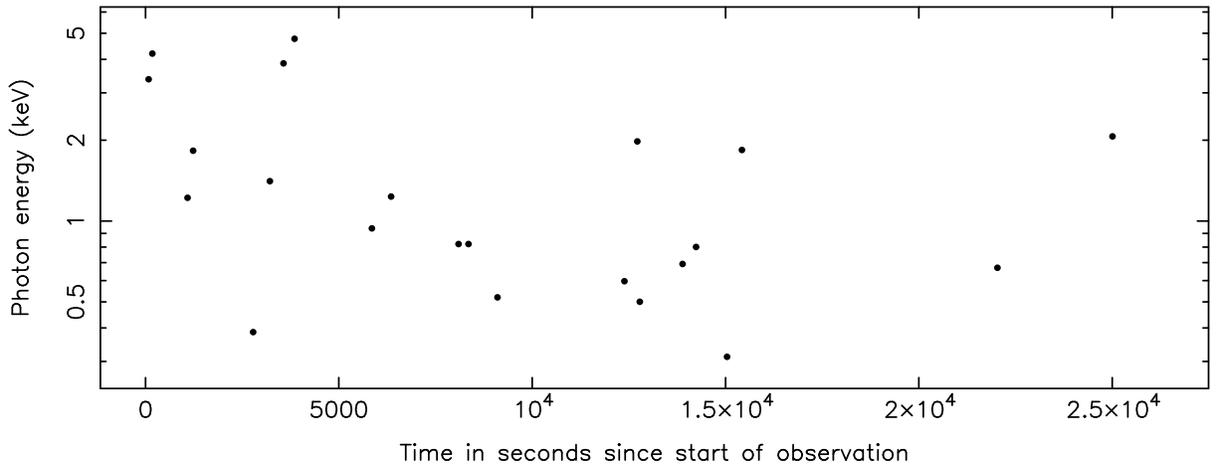}
\end{tabular}
\figcaption{
The energy of the detected photons versus the time of detection (since
the start of the observation).
\label{fig:energy-curve} }
\end{center}
\end{figure}

In Figure~\ref{fig:energy-curve} we plot the energies of the X-ray
photons (0.3--8 keV energy range) detected at the source position
against the arrival time of the photons (measured since the start of
the observation). From this figure it can be seen that 4 photons are
detected with energies above 3 keV (note that no photons were detected
with energies $>$5 keV) and the remainder of the photons have energies
below $\sim$2.5 keV. In the last $\sim$10 ksec of the observation only
two source photons are detected compared with twenty photons during
the first $\sim$15 ksec; this might suggest that the source exhibited
variability during our observation. To investigate if XTE J0929--314
was indeed variable during our observation, we applied
Kolmogorov-Smirnov and Cramer-Von Mises tests on the event list (as
shown in Figure~\ref{fig:energy-curve}) to attempt to disprove the
hypothesis that the source count rate is constant. Both tests showed
that XTE J0929--314 is variable for energies between 0.3 and 8 keV
(although effectively only up to 5 keV since no photons are detected
above that energy) at the 95\% confidence level. From
Figure~\ref{fig:energy-curve} it can be seen that the four photons
which have energies $>$3 keV all arrived within the first 5000 seconds
of the observation, which may be the main cause of the variability
detected in the source. Therefore, we also applied the above two tests
to the event list for photon energies between 0.3 and 2.5 keV. The
Kolmogorov-Smirnov test revealed that the 0.3--2.5 keV event list
might also be variable but only at a 90\% confidence level. The
Cramer-Von Mises test could not find evidence for variability in this
energy range. Therefore, we conclude that the variability observed in
the source is likely stronger at higher photon energies.

\subsubsection{Time-averaged X-ray spectrum \label{subsubsection:spectral_0929}}

Despite the low number of photons detected, we extracted the source
spectra (averaged over the whole duration of the observation) using
the CIAO tool 'psextract' for the energy range 0.3--8 keV, which also
created the response matrix and the ancillary response files (the
latter was also automatically corrected for the time-variable
low-energy quantum efficiency degradation of the CCD). We used a
circle with a radius of $1.5''$ as source extraction region. We fitted
the spectrum using Xspec version 11.3.0 (Arnaud 1996). The small
number of photons observed does not allow for $\chi^2$ statistics to
be used during the fits. Therefore, we fitted the data using Cash
statistics (Cash 1979). Since Cash statistics cannot be used on
background subtracted spectra, we did not subtract the background from
our data. The errors thus introduced are likely to be small since less
than 0.4 background photons are expected in the source extraction
region (see above). The quality of the fits was investigated by
generating 10,000 Monte Carlo simulations of the best-fit spectrum: if
the fit is good, roughly half the simulations should have values of
the Cash statistics which are lower than those of the data.

\begin{figure}
\begin{center}
\begin{tabular}{c}
\psfig{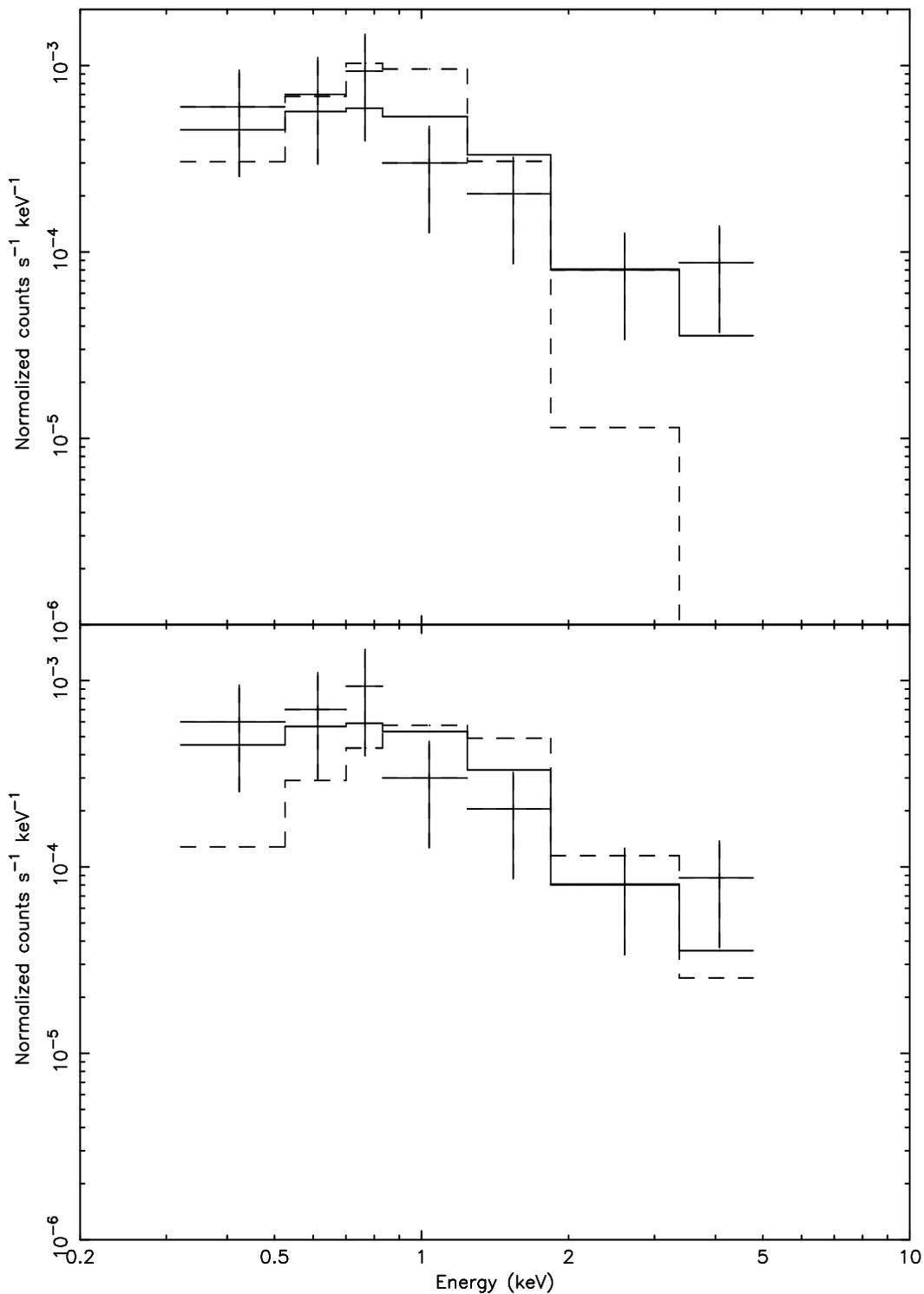}
\end{tabular}
\figcaption{
The time-averaged spectrum of XTE J0929--314. In the top panel, we
show the best power-law fit to the data (solid line) and the best NSA
fit through the data with a fixed normalization (to a distance of 10
kpc; dashed line). In the bottom panel, we again show the best
power-law fit (solid line) but now the best NSA fit with the
normalization left free (dashed line).  For display purposes, the data
were rebinned so that each bin has 3 counts but the spectra were
fitted without any rebinning.
\label{fig:spectra} }
\end{center}
\end{figure}

In all our spectral fits we left the interstellar column density
$N_{\rm H}$ as a free parameter. To calculate the errors on the
obtained fluxes, we fixed each free fit parameter one at a time,
either to its minimum or maximum allowed value as obtained from the
fits. After that we refitted the data and recalculated the
fluxes. This process was then repeated for each free parameter and the
final flux range determined the flux errors. The spectra of
neutron-star X-ray transients in their quiescent states are usually
dominated by a soft thermal component which can adequately be fitted
with a neutron-star hydrogen atmosphere model for non-magnetized stars
(the NSA model; we used that of Zavlin, Pavlov, \& Shibanov 1996; we
also assumed a mass of $1.4~M_\odot$ and a radius of 10 km for the
neutron star in XTE J0929--314). In such a model the normalization is
given by $1\over d^2$, with $d$ the distance to the source in parsecs.
The distance toward XTE J0929--314 is not known but it is constrained
to be $>$5 kpc (Galloway et al.~2002).  When leaving the normalization
free, 88\% of the simulated spectra have better Cash-statistics than
the data demonstrating that this model does not provide an accurate
description of the data (Fig.~\ref{fig:spectra} {\it bottom}).  From
this model we obtained an effective temperature $kT_\infty$ (for an
observer at infinity) of $0.3\pm0.1$ keV and the column density is
$<$$3\times 10^{20}$ cm$^{-2}$. However, the normalization was
constrained to be $<$$ 1
\times 10^{-11}$, resulting in a distance constrain of $>$316
Mpc. This unrealistic distance limit further demonstrates that a
simple NSA model does not provide a good fit to the data.

We also fitted the data using the NSA model but with the normalization
fixed and assuming three distances: 5, 10, or 15 kpc.  The results are
listed in Table~\ref{tab:spectral_fits} and shown in
Figure~\ref{fig:spectra} (top). From this table it can be seen that
such a model does not provide an acceptable fit to the data since
100\% of the simulated spectra have lower values for the
Cash-statistics than the data themselves.  We also fitted a simple
blackbody model to the data but again such a model did not provide an
adequate fit to the data with a fit quality of 0.94 and a $kT$ of
$0.6^{+0.2}_{-0.1}$ keV (and $N_{\rm H}<3 \times 10^{20}$
cm$^{-2}$). The radius of the emitting area was $< 0.12 \times {d\over
{\rm 10~kpc}}$ km, with $d$ the distance in kpc, much smaller than the
expected radius of a neutron star.  The inability of the thermal
models to provide an adequate fit to the data and the fact that four
photons are detected with energies above 3 keV point to a X-ray
spectrum which is significantly harder than a simple thermal
model. Therefore, we fitted the X-ray spectrum with a power-law model
(Fig.~\ref{fig:spectra}; Tab.~\ref{tab:spectral_fits}). The fit
quality of 0.68 using such a model is considerably better than for a
thermal model (0.68 versus $>$0.88). The fit resulted in a power-law
photon index $\Gamma$ of $1.8^{+0.6}_{-0.5}$, a column density of $<6
\times 10^{20}$ cm$^{-2}$, and a 0.5--10 keV flux of $6^{+4}_{-2}
\times 10^{-15}$ ergs s$^{-1}$ cm$^{-2}$ (corrected for
absorption). This gives an unabsorbed 0.5--10 keV X-ray luminosity of
$7^{+5}_{-2}
\times 10^{31} ({d \over {\rm 10~kpc}})^2$ ergs s$^{-1}$, with $d$ the
distance in kpc.

To obtain limits on the temperature and luminosity of a possible
thermal component in the data, we fitted the spectrum with a power-law
component plus a neutron-star atmosphere component. We again fixed the
normalization of the neutron-star atmosphere component to correspond
to 5, 10, or 15 kpc but left the other parameters free (i.e., the
column density, the effective temperature, the photon index, and the
normalization of the power-law component). The maximum allowed
temperatures and bolometric luminosities are listed in
Table~\ref{tab:nsa_limits}, together with the minimum fraction of the
0.5--10 keV flux which is due to the power-law component. The table
shows that the maximum allowed $kT_\infty$ is between 0.04 and 0.05
keV with a corresponding maximum allowed bolometric luminosity of
$0.4-1.7 \times 10^{32}$ ergs s$^{-1}$ (for both parameters the
highest values are obtained for the cases which assume the largest
distance). For all assumed distances, at least $\sim$70\% of the flux
in the 0.5--10 keV range is due to the power-law component. This
demonstrates that the power-law component dominates the X-ray emission
in the range 0.5--10 keV for XTE J0929--314.

\subsubsection{Spectral variability}

\begin{figure}
\begin{center}
\begin{tabular}{c}
\psfig{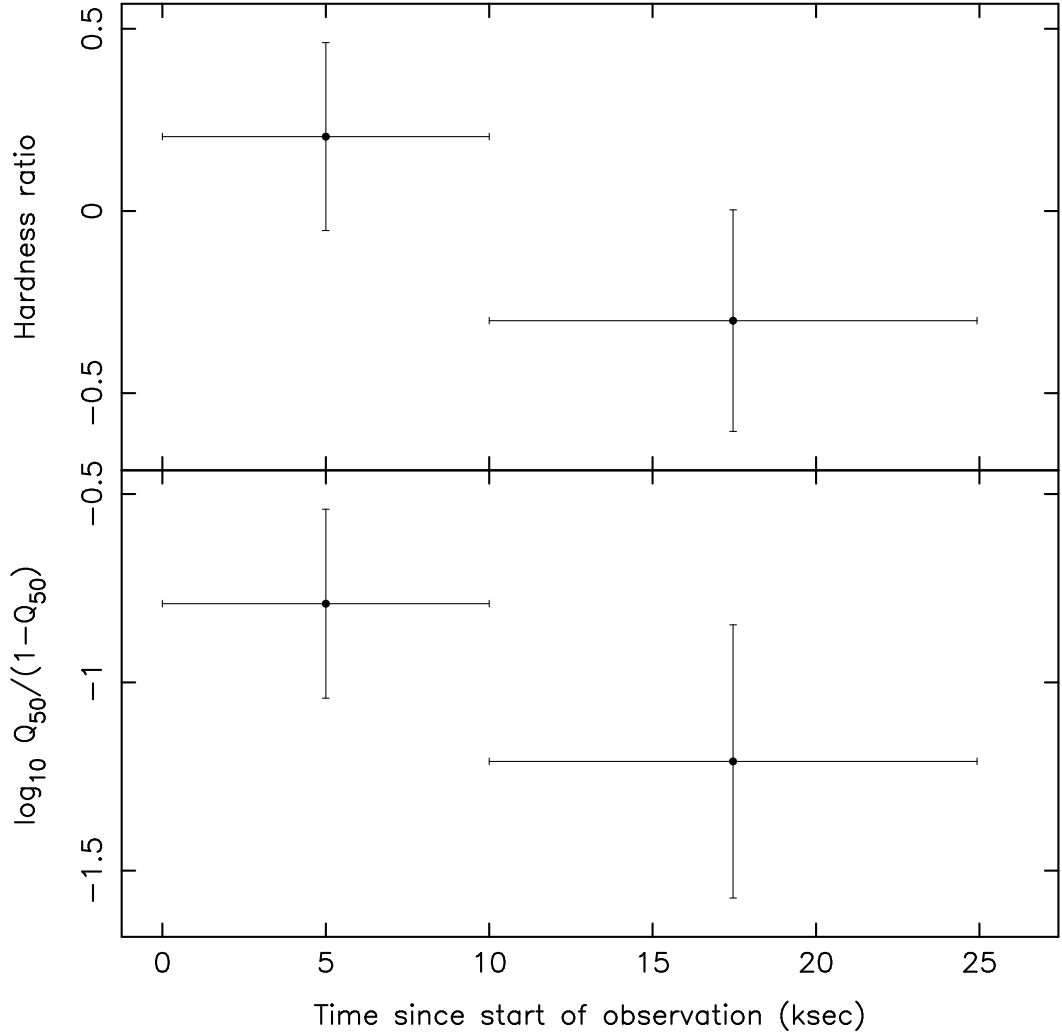}
\end{tabular}
\figcaption{
The hardness ratio ({\it top}) and the $\log{}_{10} {Q_{50} \over 1 -
Q_{50}}$ ({\it bottom}) versus the time during the observation. The
hardness ratio is defined as the logarithm of the ratio between the
number of photons detected with energies $>$1 keV and the number
detected with energies $<$1 keV, and $Q_{50}$ is the median quantile
as defined by Hong et al.~(2004).
\label{fig:quantile} }
\end{center}
\end{figure}

In \S~\ref{subsubsection:cr_0929} we demonstrated that XTE J0929--314
exhibited variability in its count rate at the 95\% confidence
level. It is possible that this count rate variability is accompanied
by or even due to changes in the X-ray spectrum of the source. If
true, the spectral results we obtained in
\S~\ref{subsubsection:spectral_0929} only represent a time-averaged
view of the spectral properties of the source. To investigate if
indeed XTE J0929--314 exhibited spectral variability we calculated
X-ray hardness ratios from the first part of the observation (between
0 and 10 ksec after the start of the observation; 13 photons were
detected in this time interval) and the second part of the observation
(the remaining $\sim$14 ksec of the observation; 9 photons were
detected). As X-ray hardness ratio we used the logarithm of the ratio
between the number of photons detected with energies $>$1 keV and the
number of photons detected with energies $<$1 keV. In
Figure~\ref{fig:quantile} ({\it top panel}) we show the hardness ratio
versus time. During the second part of the observation the hardness
ratio is smaller than during the first part, but the error bars are
large and the data are consistent with a non-variable spectrum.

Hong, Schlegel, \& Grindlay (2004) have argued that the X-ray hardness
ratios normally used in the literature (e.g., those we used in the
previous paragraph) are not appropriate for sources with low numbers
of counts. Since we have roughly 10 counts to construct our hardness
ratios, these ratios might not represent an accurate description of
the spectral shape of XTE J0929--314.  Therefore, we also calculate
so-called ``quantiles'' using the quantile method outlined by Hong et
al.~(2004)\footnote{The errors on the quantiles were calculated using
the method of Maritz \& Jarret (1978; see Hong et al.~2004) but Hong
et al.~(2004) demonstrated, using simulations, that the errors on the
quantiles are overestimated when the source counts are very low, as in
our case.  Therefore, we corrected the quantile errors using the
corrections provided by Hong et al.~(2004).}. Hong et al.~(2004) found
that the quantity $log_{10} {Q_{50}\over 1 - Q_{50}}$, with $Q_{50}$
the median quantile, provided a good indication for the spectral
hardness of a source. Therefore, in the bottom panel of
Figure~\ref{fig:quantile} we plot this quantity as a function of
time. Similar to the hardness ratios, the error bars on the quantiles
are such that a constant spectral shape throughout the observation
cannot be excluded.

\begin{figure}
\begin{center}
\begin{tabular}{c}
\psfig{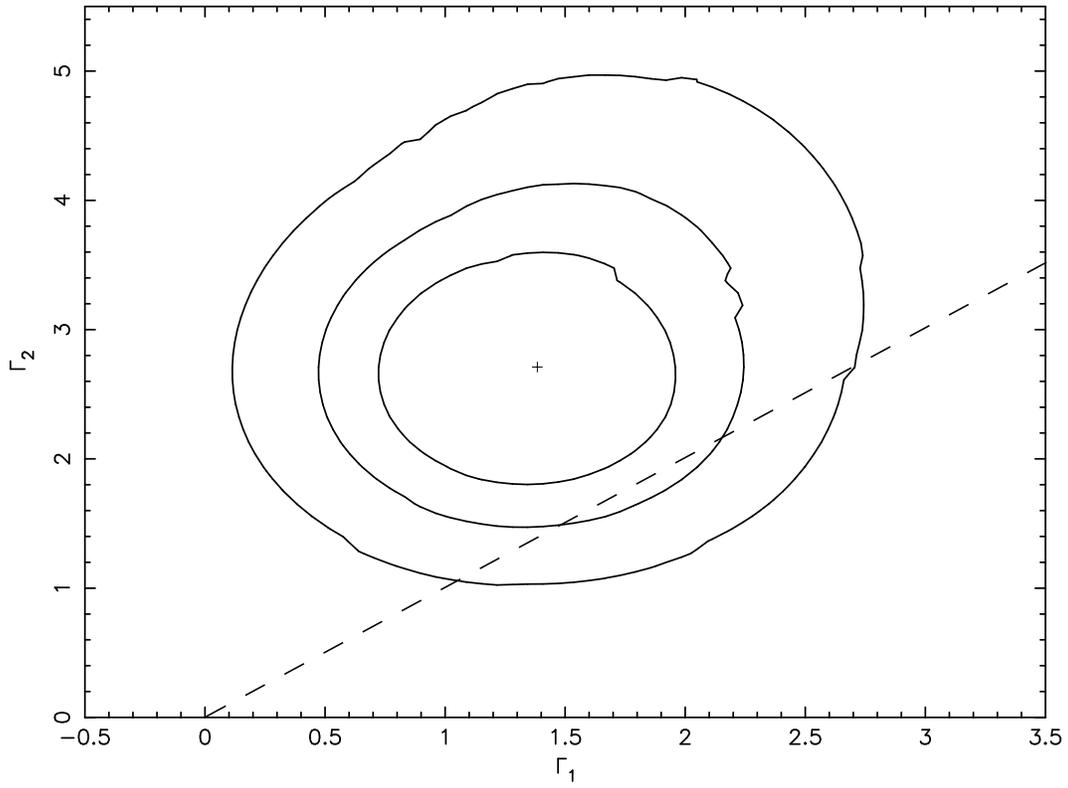}
\end{tabular}
\figcaption{
A comparison of the photon index of the first part of the data (0--10
ksec; $\Gamma_1$) with that of the second part of the data (the
remaining $\sim$14 ksec of the observation; $\Gamma_2$). The contours
are the 68\%, 90\% and 99\% confidence levels. The cross marks the
position of the best-fit values of the indices. The dashed line
indicates where both indices are equal.\label{fig:contour} }
\end{center}
\end{figure}

As a last way of investigating possible spectral variability, we
extracted the X-ray spectrum for each of the two time intervals used
above. The results of the spectral fitting are listed in
Table~\ref{tab:variability}. When fitting the two data sets
separately, a power-law model produces acceptable fits (with fit
quality of 0.6--0.7) with photon indices of 1.3$^{+0.9}_{-0.7}$ and
2.6$^{+1.2}_{-0.9}$ for the first and second part of the observation,
respectively. The values again indicate that the source spectrum had
softened during the course of the observation, although the indices
are consistent with each other (within the errors). The 0.5--10 keV
X-ray flux decreased by a factor of 3--4 during the observation indeed
suggesting variability in the brightness of the source (note, again
the errors are large and only minimal variability might have been
present). To further investigate how significant the softening of the
X-ray spectrum is, we fitted the two data sets simultaneously. During
this fit, we assumed that the column density did not vary during the
observation and we tied it between the two data sets. The fit results
were consistent with those obtained when fitting the data sets
separately (Tab.~\ref{tab:variability}). In Figure~\ref{fig:contour}
we plot the confidence regions of the photon indices, showing that the
two indices are only different from each other at the $<$90\%
confidence level. This is illustrated in the figure by the line of
equal index between two data sets which intersects the 90\% confidence
contour. We conclude that the limited statistics does not allow for
any strong conclusions on the possible softening of the X-ray spectrum
suggested by Figure~\ref{fig:energy-curve}.

We investigated if a thermal model could fit the two data sets, but
when using a NSA model with fixed normalization (to a distance of 5,
10, or 15 kpc), the fit qualities were always 1.00 for the first part
of the observation and between 0.89 and 0.94 (larger for smaller
assumed distance) for the second part of the observation. Also a
simple black-body model does not produce a good fit with fit qualities
of 0.85 and 0.87 for the first and second part of the observation,
respectively.  This clearly demonstrates that during the whole
observation, the source had an X-ray spectral shape which was
significantly different (i.e., harder) from a simple thermal
model. This confirms our conclusions in
\S~\ref{subsubsection:spectral_0929}. Investigations into whether or
not the fractional contribution of the power-law component to the
0.5--10 keV flux changed during the observation did not result in
useful insights due to the limitation on the statistics of the
individual data sets.

\subsection{XTE J1751--305 \label{subsection:1751}}

\begin{figure}
\begin{center}
\begin{tabular}{c}
\psfig{figure=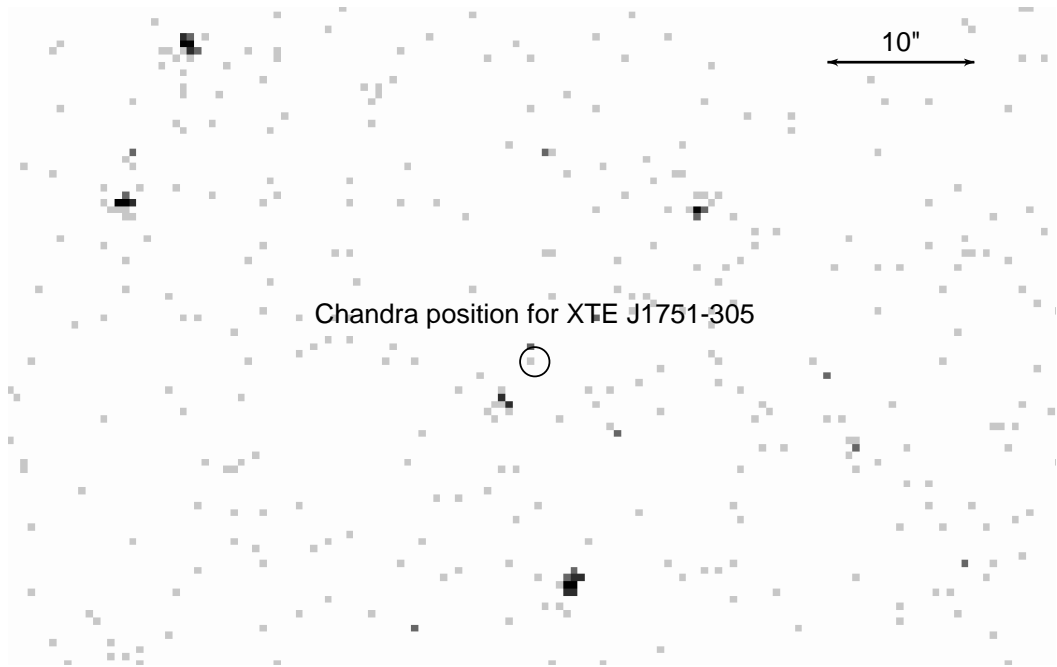,width=14cm}
\end{tabular}
\figcaption{
The 0.5--7 keV image of the region around XTE J1751--305. East is to
the left and north is up. Several point sources are detected close to
the position of XTE J1751--305 but none of these sources is consistent
with the position of XTE J1751--305 (as indicated by the circle which
has a radius of 1$''$, which means that 90\% of the energy of the
photons is encircled in this region).
\label{fig:image_1751} }
\end{center}
\end{figure}

We begin a similar analysis for XTE J1751--305 by searching for point
sources in our data (using 'wavdetect'). We detected 34 sources in our
image but XTE J1751--305 itself was not detected (a discussion of the
properties of these 34 sources, unrelated to XTE J1751--305, is beyond
the scope of this article). In Figure~\ref{fig:image_1751} we show the
0.5--7 keV X-ray image near the position of XTE J1751--305. Several
point sources are detected close to the position of XTE J1751--305
(obtained during outburst using {\it Chandra}; Markwardt et al.~2002),
but only 1 photon was detected (in the energy range 0.5--7 keV) in a
circle with a radius of 1$''$ around the position of XTE J1751--305
(for {\it Chandra}'s point-spread-function, 1$''$ corresponds to the
encirclement of 90\% of the energy), which does not constitute a
significant detection. When using also the data at lower energies
(i.e., using the energy range 0.3--7.0 keV), we still could only
detect 2 photons within an 1$''$ radius. When using a larger radius of
1.5$''$, 3 photons are detected for the energy range 0.5--7 keV, and 5
photons for the energy range 0.3--7 keV. However, in these cases more
than half the possible source photons would fall outside the circle
which should encircle $>$90\% of the energy, which is
unlikely. Moreover, the positions of the two softest photons (those
with energies $<$0.45 keV) were consistent with the position of a
bright $K$ star close to the position of XTE J1751--305 (1.2$''$ away;
the two photons have a distance of 0.5$''$ and 0.8$''$ from the
position of this star). Thus these two photons could have possibly
originated from this star and not from XTE J1751--305, especially
because the high column density toward the pulsar ($\sim$$10^{22}$
cm$^{-2}$; Miller et al.~2003) would make unlikely any detection of
its photons which have energies below 0.5 keV.

The closest, significantly detected X-ray source (the one to the
south-east of XTE J1751--305) is located 3.8$''$ away\footnote{The
coordinates of this source are: right ascension $17^h ~51^m
~13^s.673$, declination $-30\degr ~37' ~26''.40$. The coordinates are
for epoch J2000.0 and have errors of $0.6''$ (90\% confidence
levels).} from XTE J1751--305. The absolute positional accuracy of
{\it Chandra} is 0.6$''$ (90\% confidence level; $0.8''$ for 99\%
confidence level\footnote{See
http://cxc.harvard.edu/cal/ASPECT/celmon/}) strongly suggesting that
this south-east source is not the quiescent counterpart of XTE
J1751--305. It might be possible that during our observation the
positional accuracy of {\it Chandra} was (for some reason) extremely
inaccurate. To test this hypothesis, we compared our X-ray image with
the $K$-band image of the region around XTE J1751--305 published by
Jonker et al. (2003). Of the 13 X-ray sources located in the region
covered by this $K$-band image, 8 have $K$ objects located within a
radius of 0.6$''$ (including the X-ray source located 3.8$''$ away
from XTE J1751--305 as well as the four other reasonably bright X-ray
point sources visible in Fig.~\ref{fig:image_1751}). We simulated
10,000 X-ray point sources with random positions in the field of the
$K$ image. Only $<$7\% of the simulated sources have a $K$ star (of
similar or higher brightness than the potential $K$ counterparts of
the detected X-ray point sources) falling within a 0.6$''$ radius of
their positions. The coincidence rate between the detected X-ray
sources and the $K$ stars is significantly higher ($\sim$60\%) than
the simulated one demonstrating that most of the $K$ stars located in
the X-ray error circles are indeed the infrared counterparts of those
X-ray sources.  Our analysis strongly suggests that the positional
accuracy of our X-ray image is within the standard accuracy quoted for
{\it Chandra}. Therefore, we conclude that XTE J1751--305 was not
conclusively detected during our {\it Chandra} observation.

To calculate a flux upper limit for XTE J1751--305, we assume that
$<$5 photons\footnote{Using 5 photons as an upper limit can be
interpreted (Gehrels 1986) as a 97\% or 90\% confidence upper limit
for 1 or 2 detected photons from the source, respectively, which is
the number of photons we possible detected from XTE J1751--305.} (for
the energy range 0.5--7 keV) were detected from the source, resulting
in a count rate upper limit of 1.16 $\times 10^{-4}$ counts s$^{-1}$
(0.5--7 keV).  We used PIMMS\footnote{Available
http://heasarc.gsfc.nasa.gov/Tools/w3pimms.html} to calculate the flux
upper limits assuming different spectral model. As column density we
used the value as observed during outburst ($N_{\rm H} = 9.8 \times
10^{21}$ cm$^{-2}$; Miller et al.~2003) and we assume a black-body
spectral shape\footnote{Although a neutron-star atmosphere model is
usually fitted to the X-ray spectra of quiescent neutron star X-ray
transients, such a model cannot be used in PIMMS and we had to resort
to the black-body model as an approximation.} (with temperatures $kT$
of 0.05--0.7 keV) or a power-law spectral shape (with a photon index
$\Gamma$ of 0--4). Using a black-body model, we find that the 0.5--10
keV flux upper limit $F_{\rm upper}$ (in units of ergs s$^{-1}$
cm$^{-2}$) can be accurately approximated by

\begin{equation}
log_{10} F_{\rm upper} =
-14.787 + 4.09 e^{-{kT\over 0.0827~{\rm keV}}}
\label{eq:bblimit}
\end{equation}

\noindent
Using a power-law model, we find

\begin{equation}
F_{\rm upper} = 5.76\times 10^{-15} - 3.05 \times
10^{-15} \Gamma + 7.41\times 10^{-16} \Gamma^2
\label{eq:pllimit}
\end{equation}

\noindent
For a reasonable range in $kT$ of 0.1--0.3 keV (as observed in other
quiescent neutron star X-ray transients), we find 0.5--10 keV upper
limits of $0.2 - 2.7 \times 10^{-14}$ ergs s$^{-1}$ cm$^{-2}$ (the
highest upper limits are obtained when the assumed $kT$ is
smallest). For a reasonable $\Gamma$ of 1--3, the 0.5--10 keV flux
upper limit is $2.6-3.5\times 10^{-15}$ ergs s$^{-1}$ cm$^{-2}$ (the
minimum upper limit is obtained when $\Gamma \sim$2).

\section{Discussion}

We have observed the accretion-driven millisecond X-ray pulsars XTE
J0929--314 and XTE J1751--305 in their quiescent states. This triples
the number of such sources to be observed in their quiescent states
(the first one being SAX J1808.4--3658; Stella et al.~2000; Dotani,
Asai, \& Wijnands 2000; Campana et al.~2002). We could not detect XTE
J1751--305 during our observations, which resulted in upper limits on
the quiescent X-ray flux (0.5--10 keV; corrected for absorption) of
this system of $2.6-3.5\times 10^{-15}$ ergs s$^{-1}$ cm$^{-2}$ when
we assume that the source exhibited a power-law spectral shape with
photon index between 1 and 3, or $0.2 - 2.7 \times 10^{-14}$ ergs
s$^{-1}$ cm$^{-2}$ when assuming a black-body spectral shape with $kT$
of 0.1--0.3 keV.  We conclusively detected XTE J0929--314, but we
detected only 22 photons (energy range 0.3--8 keV) during our
observation which limits any detailed analysis.  The X-ray spectrum
could not be fitted with a simple thermal model alone (such as a
black-body or a neutron-star atmosphere model) but a single power-law
model provided an adequate fit to the data resulting in a 0.5--10 keV
X-ray luminosity of $7^{+5}_{-2} \times 10^{31} ({d \over {\rm
10~kpc}})^2$ ergs s$^{-1}$ (with $d$ in kpc). The upper limits on the
effective temperature of a thermal component ranged from 0.04 to 0.05
keV with corresponding upper limits on the bolometric luminosity of
0.4--1.7 $\times 10^{32}$ ergs s$^{-1}$ (both quantities are for an
observer at infinity).  We determined that at most $\sim$30\% of the
0.5--10 keV flux could have come from such a thermal component. We
could demonstrate that the source was variable at the 95\% confidence
level but, within the limitation of the statistics, we could not find
evidence for spectral variability during our observation. Next, we
will first discuss what can be inferred from the results of XTE
J0929--314 before we discuss the non-detection of XTE J1751--305.

\subsection{The power-law component of XTE J0929--314
 \label{subsection:0929_pl}}

Our results suggest that the X-ray emission of XTE J0929--314 in its
quiescent state is dominated by a power-law component and not by a
thermal component as is usually observed during the quiescent states
of neutron-star X-ray transients. This makes XTE J0929--314 similar to
SAX J1808.4--3658, which was until now the only accreting millisecond
pulsar to have been observed in its quiescent state.  In quiescence,
SAX J1808.4--3658 was found to have a 0.5--10 keV X-ray luminosity of
$\sim$$5\times10^{31}$ ergs s$^{-1}$ when fitted with a power-law
model with index of $1.5\pm0.3$ (Campana et al.~2002). This is rather
similar to the 0.5--10 keV luminosity ($\sim 7 \times 10^{31}$ ergs
s$^{-1}$ for a distance of 10 kpc) and power-law index
($1.8^{+0.6}_{-0.5}$) of XTE J0929--314. Furthermore, the quiescent
X-ray emission between 0.5 and 10 keV was in both sources dominated by
the power-law component ($>$90\% in SAX J1808.4--3658 and $>$70\% in
XTE J0929--314). Although our statistics on the XTE J0929--314 data
are poor, the only difference between the two sources seems to be that
SAX J1808.4--3658 has a somewhat lower luminosity in quiescence and
perhaps a bit harder spectrum. We note that the errors on the
parameters also suggest that both sources could have equal
luminosities and spectral hardness. Furthermore, XTE J0929--314
exhibited luminosity variability during our observation possibly
accompanied by (or even due to) spectral variability. This means that
XTE J0929--314 could be at times intrinsically fainter than SAX
J1808.4--3658 or have a harder spectrum (note that it is less likely
that XTE J0929--314 was fainter {\em and} harder than SAX
J1808.4--3658, since the indication is that the hardest spectrum was
observed when XTE J0929--314 was brightest).

The results on XTE J0929--314 and SAX J1808.4--3658 might suggest that
a power-law dominated X-ray spectrum in quiescence and a low quiescent
0.5--10 keV luminosity are common properties of accretion-driven
millisecond X-ray pulsars. However, recent work (as discussed below)
on two other weak quiescent neutron-star X-ray transients, which do
not exhibit pulsations, suggests that this might instead be a property
of many weak quiescent neutron-star X-ray transients in general and
not only of the accreting millisecond X-ray pulsars.  Jonker et
al.~(2004a) and Tomsick et al.~(2004) reported on quiescent X-ray
observations of SAX J1810.8--2609 and XTE J2123--058,
respectively. For both sources it was found that their quiescent X-ray
spectra could be fitted with a power-law model with indices of $\sim$3
and an 0.5--10 keV X-ray luminosity around $\sim$$10^{32}$ ergs
s$^{-1}$. No thermal component could be detected with a maximum
contribution to the 0.5--10 keV flux of $\sim$50\% in SAX
J1810.8--2609 (Jonker et al.~2004a) and $\sim$60\%--70\% in XTE
J2123--058 (Tomsick et al.~2004). These luminosities are somewhat
higher than those of SAX J1808.4--3658 and XTE J0929--314 and the
X-ray spectra somewhat softer, but they are among the lowest
luminosity quiescent neutron-star X-ray transients for which no X-ray
pulsations were seen during their outbursts.

Jonker et al.~(2004a, b) used the results obtained for SAX
J1808.4--3658, SAX J1810.8--2609, and XTE J2123--058 together with
those obtained for the other quiescent neutron star X-ray transients,
to show evidence that, in their quiescent spectra, the contributions
of the power-law components to the 0.5--10 keV flux seem to increase
(from at most several tens of percents to over 90\%) when the
quiescent source luminosities decrease from $\sim$$10^{33}$ ergs
s$^{-1}$ to approximately $5\times 10^{31}$ ergs s$^{-1}$. They also
found a similar trend for higher quiescent luminosities of the
transients: the fractional power-law contribution also increases when
the quiescent source luminosities increase from the comparison
luminosity of $1-2 \times 10^{33}$ ergs s$^{-1}$. Our results show
that XTE J0929--314 fits between SAX J1808.4--3658 on the
low-luminosity side, and SAX J1810.8--2609 and XTE J2123--058 on the
higher-luminosity side, which provides additional support to the
possible correlations found by Jonker et al.~(2004a, b).  This result
would suggest that hard quiescent emission is not a feature unique to
accretion-driven millisecond X-ray pulsars but rather a general
feature of quiescent neutron-star X-ray transients when they are at
low quiescent luminosities (with increasingly harder spectra when the
luminosities get lower).  The current data still suggest that the
accretion-driven millisecond X-ray pulsars might have slightly lower
luminosities and harder spectra than the other sources, but it remains
to be seen if this holds when additional neutron-star X-ray transients
are observed in their quiescent states. Additional neutron star X-ray
transients need to be studied in their quiescent states to better
understand the power-law spectral component in the quiescent
spectra. Such studies might also clarify if the quiescent transient in
Terzan 5 (EXO 1745--248; Wijnands et al.~2004), which does not follow
this correlation (Jonker et al.~2004b), is unique among the quiescent
neutron star systems, or if it is the first example of many similar
systems which would suggest that this correlation is spurious.

\subsection{The non-detection of a thermal component in XTE J0929--314}

We can investigate whether our upper limits on the contribution of the
thermal component to the quiescent spectrum of XTE J0929--314 are
consistent with what would be expected from the cooling neutron star
model proposed by Brown et al.~(1998). The quiescent thermal
luminosity predicted by this model depends on the time-averaged (over
$>$10,000 years) accretion rate of the source. For XTE J0929--314 we
can estimate its time-averaged accretion rate in two ways; either by
using the outburst fluence of the source and the limited information
about the outburst recurrence time, or by assuming that the mass
transfer is solely due to gravitational radiation and all matter is
eventually accreted onto the neutron star.

\subsubsection{Constraints obtained when using the 2002 outburst fluence
\label{subsubsection:0929_outburstfluence}}

According to Galloway et al.~(2002), the 2--60 keV fluence of the 2002
outburst of XTE J0929--314 was $4.2\times 10^{-3}$ ergs cm$^{-2}$,
with an estimated bolometric correction of $\sim$2.34. This results in
a bolometric fluence of $\sim$$9.8\times 10^{-3}$ ergs
cm$^{-2}$. During the lifetime of {\it RXTE} (which is at present 8.5
years), only 1 outburst has been observed, resulting in a time
averaged accretion flux $\langle F_{\rm acc} \rangle$ of $<3.7\times
10^{-11}$ ergs s$^{-1}$ cm$^{-2}$. Using the Brown et al.~(1998) model
and assuming standard core cooling, the expected quiescent flux
$F_{\rm q}$ is approximately given by $F_{\rm q} = {\langle F_{\rm
acc} \rangle\over 135}$ (Brown et al.~1998; Wijnands et al.~2001;
Rutledge et al.~2002). For XTE J0929--314, this results in $F_{\rm q}
< 2.7 \times 10^{-13}$ ergs s$^{-1}$ cm$^{-2}$, which is consistent
with the upper limits on the bolometric fluxes of a possible
neutron-star atmosphere component in the quiescent spectra (which are
$<$1.5 $\times 10^{-14}$ ergs s$^{-1}$ cm$^{-2}$). However, in this
scenario the source would have to remain quiescent for a significant
period (i.e., much longer than the 8.5 years we assumed here) to allow
the predicted limits to come down to the measured flux limits.

Using our measured flux limits, we can actually estimate how long XTE
J0929--314 has to be in quiescence for the neutron star surface to be
as cool as we measure if the neutron star cools down using standard
core cooling processes. From $F_{\rm q} = {\langle F_{\rm acc}
\rangle\over 135}$ it can be derived (Wijnands et al.~2001) that $F_{\rm
q} \approx {t_{\rm o} \over t_{\rm o} + t_{\rm q}} \times {\langle
F_{\rm o} \rangle \over 135} $, with $\langle F_{\rm o} \rangle$ the
averaged flux during outburst, $t_{\rm o}$ the averaged time the
source is in outburst, and $t_{\rm q}$ the averaged time the source is
in quiescence. From Galloway et al.~(2002), it follows that $t_{\rm o}
\approx 73$ days (if the 2002 outburst had a duration which is typical
for the source) and $\langle F_{\rm o} \rangle =1.6
\times 10^{-9}$ ergs s$^{-1}$ cm$^{-2}$. Using our limits on $F_{\rm
q}<1.5 \times 10^{-14}$ ergs s$^{-1}$ cm$^{-2}$, then $t_{\rm q}>
5.8\times 10^4$ days, resulting in $t_{\rm q}> 160 $ years. Although
this predicted quiescent period is quite long compared to the known
recurrence time of several other neutron-star X-ray transients
(ranging from less than 1 year up to several decades; e.g., Chen,
Shrader, \& Livio 1997), the disk instability model proposed for the
outburst behavior of X-ray transients (Lasota 2001) might be able to
accommodate such long quiescent episodes. We note that due to the very
low peak fluxes during the 2002 outburst, we cannot rule out that,
prior to the launch of {\it RXTE}, one or more outbursts occurred
during the last four decades (i.e., since the birth of X-ray
astronomy). Weak outbursts, such as the 2002 outburst of XTE
J0929--314, could have easily been missed by X-ray satellites other
than {\it RXTE}.

\subsubsection{Constraints obtained when assuming mass 
transfer solely via gravitational radiation}

Galloway et al.~(2002) stated that if the mass transfer in XTE
J0929--314 is driven by gravitational radiation then the time-averaged
mass transfer rate $\dot{M}_{\rm GR}$ in this system should be

\begin{equation}
\dot{M}_{\rm GR} = 5.5 \times 10^{-12} ({ M_{\rm ns} \over
 1.4~M_\odot})^{2\over 3} ({M_{\rm c} \over 0.01~M_\odot})^2
 ~M_\odot~{\rm yr}^{-1} \label{equation:gw}
\end{equation}

\noindent
with $M_{\rm ns}$ the mass of the neutron star and $M_{\rm c}$ the
mass of the companion star (both in solar masses).  From the model by
Brown et al.~(1998; see also Rutledge et al.~2002) the quiescent
luminosity $L_{\rm q}$ can be estimated via

\begin{equation}
L_{\rm q} = 8.7 \times 10^{33} ( {\langle \dot{M} \rangle \over
10^{-10} M_\odot~{\rm yr}^{-1}} ) {Q \over 1.45~{\rm MeV}} {\rm
~ergs~s}^{-1} \label{equation:lq}
\end{equation}

\noindent
with $\langle \dot{M} \rangle$ the time-averaged mass accretion rate
onto the neutron star, and $Q$ the amount of heat deposited in the
crust per accreted nucleon (e.g., Haensel \& Zdunik 1990). If we
assume that the matter transfer from the companion star is solely
driven by gravitational radiation and that all the matter transferred
from the companion star is eventually accreted onto the neutron star
then $\langle \dot{M} \rangle = \dot{M}_{\rm GR}$. Substituting
equation~\ref{equation:gw} in equation~\ref{equation:lq} gives

\begin{equation}
L_{\rm q} = 4.8 \times 10^{32} ({ M_{\rm ns} \over 1.4
M_\odot})^{2\over 3} ({M_{\rm c} \over 0.01 M_\odot})^2 {Q \over
1.45~{\rm MeV}} {\rm ~ergs~s}^{-1}
\end{equation}

\noindent
If $ M_{\rm ns} =1.4~M_\odot$, then Galloway et al.~(2002) obtained a
minimum mass for the companion star of $0.008~M_\odot$ (thus $M_{\rm
c}>0.008~M_\odot$). If we further assume $Q= 1.45$ MeV, then $L_{\rm
q} > 3 \times 10^{32}$ ergs s$^{-1}$. Only for large distances ($>$15
kpc) can the upper-limits on any contribution of a thermal component
be consistent with this lower-limit on the quiescent luminosity (if
$M_{\rm c}$ is not much greater than the minimum allowed value of
0.008 $M_\odot$). For smaller distances the predicted quiescent
luminosity for a thermal component is a factor 2--8 times too high
(this discrepancy becomes worse if $M_{\rm c}$ becomes larger). Firm
conclusions can only be obtained when the exact luminosity of the
thermal component can be measured with better data. However, our
results already suggest that the quiescent luminosity predicted is
higher than that measured if we assume that the matter transfer from
the companion star is solely driven by gravitational radiation and
that all the matter transferred is accreted onto the neutron
star. Possible reasons for this discrepancy are that not all the
matter transferred is accreted onto the neutron star or that enhanced
cooling processes occur in the neutron-star core.

\subsection{The non-detection of XTE J1751--305}

Despite the fact that we could not conclusively detect XTE J1751--305
during our observation, the limits on the X-ray flux of this source
can still be used to provide further insight into the quiescent
emission of neutron star X-ray transients. We do not know the spectral
shape of the quiescent emission of this system and therefore in
\S~\ref{subsection:1751} we calculated the flux upper limits assuming
two different spectral models (either a black-body or a power-law
model; equations~\ref{eq:bblimit} and~\ref{eq:pllimit}). The limits
obtained using the black-body model are most useful for constraining
the cooling neutron star model which assumes that the emission is due
to thermal radiation from the neutron-star surface. We will attempt to
constrain this model once again taking into account the limited
observational knowledge about the time-averaged accretion rate of the
source (\S~\ref{subsubsection:1751_fluence}) and assuming that the
mass transfer is solely due to gravitational radiation and that all
matter will eventually be accreted by the neutron star
(\S~\ref{subsubsection:1751_gr}) . Here we first discuss the
possibility that the source emitted a power-law shaped quiescent X-ray
spectrum.

The two other accretion-driven millisecond X-ray pulsar which have
been observed in quiescence have shown a power-law dominated quiescent
X-ray spectrum with photon indices of 1.5$\pm0.3$ (SAX J1808.4--3658;
Campana et al.~2002) and 1.8$\pm0.6$ (XTE J0929--314;
Tab.~\ref{tab:spectral_fits}). If we assume that XTE J1751--305 has
also a power-law shaped quiescent spectrum, then, using
equation~\ref{eq:pllimit}, we get 0.5--10 keV flux upper limits on the
quiescent emission of $2.5- 3.5 \times 10^{-15}$ ergs s$^{-1}$
cm$^{-2}$ (assuming a range of 1--3 for the photon index, which covers
the allowed values of the photon indices for SAX J1808.4--3658 and XTE
J0929--314), giving 0.5--10 keV luminosity upper limits of $1.9 - 2.7
\times 10^{31} ({d\over 8 {\rm ~kpc}})^2$ ergs s$^{-1}$ ($d$ in
kpc). If the distance toward XTE J1751--305 is $<$11 kpc (thus the
source is located in the Galactic center region) and the quiescent
spectrum follows a power-law shape, then XTE J1751--305 would be the
intrinsically faintest quiescent neutron-star X-ray transients so far
known. However, if the source is located at a larger distance, SAX
J1808.4--3658 would still remain the faintest quiescent system (with a
luminosity of $\sim 5 \times 10^{31}$ ergs s$^{-1}$; Campana et
al.~2002), unless it can be confirmed that XTE J0929--314 can become
at times fainter than SAX J1808.4--3658 (see
\S~\ref{subsection:0929_pl}).

\subsubsection{Constraints obtained when using the 2002 outburst fluence 
\label{subsubsection:1751_fluence}}

According to Markwardt et al.~(2002), the 2--200 keV fluence of the
2002 outburst of XTE J1751--305 was $2.5 \times 10^{-3}$ ergs
cm$^{-2}$. Since no bolometric correction factor is currently
available, we will use this fluence, but we note that the bolometric
fluence could be significantly higher (a factor of a few) and thus
also the predicted quiescent thermal flux. Markwardt et al.~(2002)
found a previous outburst of the source in June 1998, resulting in a
recurrence time of $\sim$3.8 years. If this is the typical recurrence
time of the source, then the time-averaged accretion rate will be
$\langle F_{\rm acc} \rangle \sim 2.1 \times 10^{-11}$ ergs s$^{-1}$
cm$^{-2}$. Since the predicted quiescent flux is given by $F_{\rm q} =
{\langle F_{\rm acc} \rangle\over 135}$
(\S~\ref{subsubsection:0929_outburstfluence}), this results in $F_{\rm
q} = 1.6 \times 10^{-13}$ erg s$^{-1}$ cm$^{-2}$, which is
significantly larger than the maximum bolometric flux ($2.7
\times 10^{-14}$ ergs s$^{-1}$ cm$^{-2}$) we obtained for a possible
thermally shaped quiescent spectrum. It might be possible that the
average recurrence time of the source is larger than the 3.8 years
found for the two outbursts so far observed. Using the method outlined
in \S~\ref{subsubsection:0929_outburstfluence}, we find that the
source must have typical quiescent episodes of $>$22 years for the
predicted quiescent flux to come down to the calculated flux
limits. Such a recurrence time scale is not uncommon among the
neutron-star X-ray transients. Another possibility for the discrepancy
between the predicted quiescent flux and the limits on the actual flux
is that the neutron star in XTE J1751--305 is colder than expected due
to enhanced cooling in its core.

\subsubsection{Constrains obtained when assuming mass transfer solely 
via gravitational radiation \label{subsubsection:1751_gr}}

Similar to XTE J0929--314, it also has been suggested for XTE
J1751--305 (Markwardt et al.~2002) that the mass transfer might be due
to gravitational radiation. If true, then $\dot{M}_{\rm GR}$ in this
system should be (Markwardt et al.~2002)

\begin{equation}
\dot{M}_{\rm GR} = 1.2 \times 10^{-11} ({ M_{\rm ns} \over
 1.4~M_\odot})^{2\over 3} ({M_{\rm c} \over 0.0137~M_\odot})^2
 ~M_\odot~{\rm yr}^{-1}. \label{equation:gw2}
\end{equation}

\noindent
If we assume that the matter transfer is solely driven by
gravitational radiation and that all the matter is eventually accreted
onto the neutron star then we can again set $\langle \dot{M} \rangle =
\dot{M}_{\rm GR}$. Substituting equation~\ref{equation:gw2} in
equation~\ref{equation:lq} gives for the predicted quiescent
luminosity (again assuming $Q = 1.45$ MeV)

\begin{equation}
L_{\rm q} = 1.0 \times 10^{33} ({ M_{\rm ns} \over 1.4
M_\odot})^{2\over 3} ({M_{\rm c} \over 0.0137 M_\odot})^2 {\rm
~ergs~s}^{-1}.
\end{equation}
 
\noindent
Markwardt et al~(2002) obtained a minimum mass for the companion star
of $M_{\rm c} > 0.0137~M_\odot$, when assuming a neutron star mass of
$1.4~M_\odot$. Using these values for the masses of the companion and
the neutron star, we find that the predicted quiescent luminosity is
given by $L_{\rm q} > 1\times 10^{33}$ ergs s$^{-1}$. The 0.5--10 keV
flux upper limit assuming a black body spectral model is given by
equation~\ref{eq:bblimit}. If $kT> 0.1$ keV (which is typically
observed for other neutron star X-ray transients in their quiescent
states), this gives a 0.5--10 keV luminosity upper limit of $>2\times
10^{32} {d \over ~8~{\rm~kpc}}$ ergs s$^{-1}$ ($d$ in kpc). The
bolometric luminosity limit could be a factor of a few larger. Only
for relatively large distances ($>10-15$ kpc; depending on the
bolometric correction factor) can the measured upper limit on the
luminosity become consistent with the predicted luminosity. For
effective temperatures below 0.1 keV, the bolometric luminosity as
calculated via $L_{\rm bol} = 4\pi
\sigma R^2 T_{\rm eff}^4$ (with $R$ and $T$ the neutron star radius
and surface temperature for an observer at infinity and $\sigma$ the
Stefan-Boltzmann constant) will become quite low and will not be
consistent with the predicted luminosity. This indicates that the
effective temperature cannot be too low in order to be consistent with
the predicted quiescent luminosity.

\subsection{Constraining the magnetic field of the neutron stars in 
XTE J0929--314 and XTE J1751--305}

Recently, Burderi et al.~(2002) and Di Salvo \& Burderi (2003)
constrained the magnetic field strengths of the neutron stars in
several neutron-star X-ray transients (i.e., KS 1731--260, SAX
J1808.4--3658 and Aql X-1) based on their measured quiescent
luminosities and on the knowledge of the spin rate of their neutron
stars\footnote{SAX J1808.4--3658 exhibits coherent oscillations in its
persistent X-ray emission during outbursts which reflect the spin rate
of the neutron star (Wijnands \& van der Klis 1998). KS 1731--260 and
Aql X-1 exhibit nearly-coherent oscillations during thermonuclear
flashes on the neutron star surfaces (Smith, Morgan, \& Bradt 1997;
Zhang et al.~1998). Those oscillations are likely to be directly
related to the neutron star spin rate as well (see, e.g., Strohmayer
\& Bildsten 2004 for a recent review).}. To constrain the magnetic
field strength, they used two proposed mechanisms which might produce
X-rays in quiescence: {\it a}) residual accretion onto the neutron
star at a very low accretion rate; {\it b}) neutron star rotational
energy which is converted into radiation (a fraction of which might be
released in X-rays) because of the presence of a rotating magnetic
dipole. They note that thermal emission from a cooling neutron star
which is heated during outburst might also contribute to the quiescent
X-ray emission and that the measured quiescent luminosities are
therefore an upper limit on the X-ray contribution due to the above
discussed scenarios.

They divided scenario {\it a} into two possibilities. The first ({\it
a1}) assumes that the radius of the magnetosphere is inside the
co-rotation radius and accretion onto the neutron star surface is
possible; the second ({\it a2}) assumes that the magneto-spheric
radius is outside the co-rotation radius but inside the radius of the
light cylinder and accretion onto the neutron star is not possible but
an X-ray emitting accretion disk might still exist with an inner
radius larger than the co-rotation radius. Scenario {\it b} can also
be divided into two possibilities. The first ({\it b1}) assumes that
the X-ray emission in quiescence is due to reprocessing of part of the
bolometric luminosity from the rotating neutron star into X-rays in
the shock front between the pulsar wind and the circumstellar matter;
the second ({\it b2}) assumes that the X-ray emission is intrinsic
emission from the radio pulsar.
 
Following the method outlined by Burderi et al.~(2002) and Di Salvo \&
Burderi (2003) and using the maximum quiescent luminosity from our
fits ($1.2\times 10^{32} ({d \over {\rm 10~kpc}})^2$ ergs s$^{-1}$,
with $d$ in kpc) and the spin rate (185 Hz; Galloway et al.~2002) of
XTE J0929--314, we found that the magnetic field strength of the
neutron star in this system should be $<2\times 10^7 {d \over {\rm
10~kpc}}$ Gauss (scenario {\it a1}), $<3\times 10^{9} {d \over {\rm
10~kpc}}$ Gauss ({\it a2}), $<5 \times 10^8 {d \over {\rm 10~kpc}}$
Gauss ({\it b1}), and $<3\times 10^9 ({d \over {\rm 10~kpc}})^{0.76}$
Gauss ({\it b2}). Therefore, from these four scenarios the
magnetic field strength of the neutron star in XTE J0929--314 can be
constrained to be $<3\times 10^9 {d \over {\rm 10~kpc}}$ Gauss.

We also calculated the limits on the magnetic field strength of the
neutron star in XTE J1751--305. The maximum flux limit of $2.7\times
10^{-14}$ ergs s$^{-1}$ cm$^{-2}$ we obtained was for a spectral model
which we assumed to be a simple black body shape, resulting in a
luminosity limit of $2\times 10^{32} {d \over 8 {\rm ~kpc}}$ ergs
s$^{-1}$ (with $d$ in kpc). Using the spin frequency of 435 Hz
(Markwardt et al.~2002), this gives limits on the magnetic field
strength of $<1\times 10^7 {d \over 8 {\rm ~kpc}}$ Gauss (scenario
$a1$), $<5 \times 10^8 {d \over 8 {\rm ~kpc}}$ Gauss ($a2$), $<1
\times 10^8 {d \over  8 {\rm ~kpc}}$ Gauss ($b1$),  and $7 \times 10^8
{d \over 8 {\rm ~kpc}}$ Gauss ($b2$).  When assuming a power-law
spectral model for the quiescent spectrum of XTE J1751--305, the upper
limits on the fluxes are significantly smaller, with a maximum
luminosity limit of $3\times 10^{31} {d \over 8 {\rm ~kpc}}$ (with $d$
in kpc), which would make the magnetic field limits a factor of $>$2
less (the maximum upper limit would be again for scenario $b2$ and is
$<3 \times 10^8 {d \over 8 {\rm ~kpc}}$ Gauss).  Therefore, from these
four scenarios the magnetic field strength of the neutron star in XTE
J1751--305 can be constrained to be $< 3- 7 \times 10^8 {d \over {\rm
8~kpc}}$ Gauss (depending on assumed spectral model for the quiescent
spectrum).

Although this method still has significant uncertainties (i.e., due to
the lack of understanding of the structure of the accretion flow
geometry in quiescent X-ray transients; see Burderi et al.~2002 and Di
Salvo \& Burderi 2003 for a more detailed discussion), our limits on
the magnetic field of the neutron stars in XTE J0929--314 and XTE
J1751--305 are consistent with the expectation that these neutron
stars have a low but non-negligible (since they exhibit pulsations
during accretion outbursts) magnetic field strength.

\acknowledgements

We thank Peter Jonker for allowing us to use his $K$-band image of the
region around XTE J1751--305 and for comments on an early version of
this paper. We thank Jae Sub Hong for useful discussion about the
quantile analysis method.

\clearpage

\clearpage

\clearpage

\clearpage

\clearpage

\clearpage

\clearpage

\begin{deluxetable}{lccccc}
\tablecolumns{6}
\tablewidth{0pt} 
\tablecaption{Spectral fits of XTE J0929--314 \label{tab:spectral_fits}}
\tablehead{
Model$^a$                         & $N_{\rm H}$             & $kT_\infty$                 & $\Gamma$            & $F^b$                                & Fit quality$^c$   \\
                                  &($10^{22}$ cm$^{-2}$)    & (keV)                       &                     & ($10^{-15}$ ergs s$^{-1}$ cm$^{-2}$) & 
}
\startdata
NSA ($d$ = 5 kpc)                 &  $0.3^{+0.2}_{-0.1}$    &  0.045$\pm0.005$            &  ---                & 9$^{+7}_{-4}$                        & 1.00 \\ 
NSA ($d$ = 10 kpc)                &  $0.2\pm0.2$            &  0.058$^{+0.008}_{-0.009}$  &  ---                & 9$^{+9}_{-5}$                        & 1.00 \\
NSA ($d$ = 15 kpc)                &  $0.2^{+0.2}_{-0.1}$    &  0.067$\pm0.08$             &  ---                & 9$^{+7}_{-4}$                        & 1.00  \\
Power-law                         &  $<$0.06                &   ---                       & $1.8^{+0.6}_{-0.5}$ & 6$^{+4}_{-2}$                        & 0.68  \\
\enddata

\tablenotetext{\,}{$^a$ For the neutron-star hydrogen atmosphere (NSA) model we used that of
Zavlin et al.~(1996) with a mass of 1.4 $M_\odot$, a radius of 10 km,
and a zero magnetic field strength for the neutron star in XTE
J0929--314.  The errors on the fit parameters are for 90\% confidence
levels.}

\tablenotetext{\,}{$^b$ The 0.5--10 keV unabsorbed flux.}

\tablenotetext{\,}{$^c$ The fit quality represents the fraction of 10,000 simulated spectra
which have Cash-statistics better than that obtained for the actual
data.}

\end{deluxetable}

\clearpage

\begin{deluxetable}{lccccc}
\tablecolumns{6}
\tablewidth{0pt} 
\tablecaption{Limits on any NSA contribution  in XTE J0929--314$^a$ \label{tab:nsa_limits}}
\tablehead{
Distance $d$ & $N_{\rm H}$ & $\Gamma$            & $kT_\infty$ & Bolometric luminosity$^b$ & Fraction in the     \\
(kpc)        & ($10^{22}$ cm$^{-2}$) &                     & (keV)       & ($10^{32}$ ergs s$^{-1}$) & power-law component$^c$ }
\startdata
5            & $<$0.14     & 1.2$^{+0.9}_{-1.1}$ & $<$0.04     & $<$0.4                    & $>$0.77 \\
10           & $<$0.10     & 1.1$^{+1.0}_{-1.2}$ & $<$0.05     & $<$1.2                    & $>$0.70 \\
15           & $<$0.08     & 1.1$^{+1.0}_{-1.3}$ & $<$0.05     & $<$1.7                    & $>$0.78 \\
\enddata

\tablenotetext{\,}{$^a$ 
Results of spectral fits assuming a NSA plus a power-law model.  For
the NSA model we used that of Zavlin et al.~(1996) with a mass of 1.4
$M_\odot$, a radius of 10 km, and a zero magnetic field strength for
the neutron star in XTE J0929--314. The upper limits are for 90\%
confidence levels. }

\tablenotetext{\,}{$^b$ Bolometric luminosity of the neutron-star 
atmosphere component as seen by an observer at infinity}

\tablenotetext{\,}{$^c$ The minimum fraction of the 0.5--10 keV flux in the power-law
component}

\end{deluxetable}

\clearpage

\begin{deluxetable}{lcccc}
\tablecolumns{5}
\tablewidth{0pt} 
\tablecaption{Possible spectral variability  in XTE J0929--314 \label{tab:variability}}
\tablehead{
Time selection     &  $N_{\rm H}$           & $\Gamma$ & Flux$^a$ & Fit quality \\ 
(ksec since start) &  ($10^{22}$ cm$^{-2}$) &          & ($10^{-15}$ ergs s$^{-1}$ cm$^{-2}$) & }
\startdata
\multicolumn{5}{c}{Separate fits}\\
\hline
0 -- 10      &   $<$0.2              & $1.3^{+0.9}_{-0.7}$ & 14$^{+12}_{-4}$                      & 0.70 \\
10 -- end    &   $<$0.1              & $2.6^{+1.2}_{-0.9}$ & 3$^{+2}_{-1}$                        & 0.62 \\
\hline
\multicolumn{5}{c}{Simultaneous fits with tied $N_{\rm H}$}\\
\hline
0 -- 10      &  $<$0.07              & $1.3\pm0.7$         & 14$^{+12}_{-6}$                      & 0.66 \\
10 -- end    &     ``                & $2.5^{+1.0}_{-0.9}$ & 3$^{+2}_{-1}$                        & `` \\
\enddata

\tablenotetext{\,}{$^a$ The 0.5--10 keV unabsorbed flux}

\end{deluxetable}


\begin{references}

\reference{}Arnaud, K. 1996, in G. Jacoby \& J. Barnes (eds.), {\it
Astronomical Data Analysis Software and Systems V.}, Vol. 101, p. 17,
ASP Conf. Series.
                             
\reference{}
Asai, K., Dotani, T., Hoshi, R., Tanaka, Y., Robinson, C. R., \&
Terada, K. 1998, \pasj, 50, 611
            
\reference{}Brown, E. F., Bildsten, L., \& Rutledge, R. E. 1998, \apj,
504, L95
    
\reference{}
Burderi, L. et al.~2002, \apj, 574, 930
    
\reference{}
Cash, W. 1979, \apj, 228, 939
    
                                     
\reference{}
Campana, S. \& Stella, L. 2000, \apj, 541, 849
                                     	 
\reference{}Campana, S., Colpi, M., Mereghetti, S., Stella, L.,
Tavani, M. 1998, \aapr, 8, 279
                                     
    
\reference{}
Campana, S., et al. 2002, \apj, 575, L15
               
\reference{}
Chen, W., Shrader, C. R., \& Livio, M. 1997, \apj, 491, 312
    
               
\reference{}
Di Salvo, T. \& Burderi, L. 2003, \aap, 397, 723
    
\reference{}
Dotani, T., Asai, K., \& Wijnands, R. 2000, \apj, 543, L145
    
\reference{}
Galloway, D. K., Chakrabarty, D., Morgan, E. H., \& Remillard,
R. A. 2002, \apj, 576, L137
    
\reference{}
Gehrels, N. 1986, \apj, 303, 336

\reference{}
Greenhill, J. G., Giles, A. B., \& Hill, K. M. 2002, \iaucirc, 7889
    
\reference{}
Haensel, P. \& Zdunik J. L. 1990, \aap, 227, 431
               
\reference{}Hong, J., Schlegel, E. M., \& Grindlay, J. E. 2004, ApJ, 
in press (astro-ph/040646)
    
\reference{}Jonker, P. G., et al.~2003, \mnras, 344, 201
    
\reference{}Jonker, P. G., Wijnands, R., \& van der Klis,
M. 2004a, \mnras, 349, 94
    
\reference{}
Jonker, P. G., Galloway, D. K., McClintock, J. E., Buxton, M., Garcia,
M., \& Murray, S. 2004b, \mnras, in press (astro-ph/0406208)
    
    
\reference{}
Juett, A. M., Galloway, D. K., \& Chakrabarty, D. 2003, \apj, 587, 754
    
\reference{}
Lasota, J.-P. 2001, NewA Rev., 45, 449
    
\reference{}
Maritz, J. S. \& Jarrett, R. G. 1978, Journal of American Statistical
Association, 73, 194
    
\reference{}
Markwardt, C.B \& Swank, J. H. 2003, \iaucirc, 8144
    
\reference{}
Markwardt, C. B., Swank, J. H., Strohmayer, T. E., in 't Zand,
J. J. M., \& Marshall, F. E. 2002, \apj, 575, L21
    
\reference{}
Markwardt, C. B., Smith, E., \& Swank, J. H. 2003, \iaucirc 8080
    
    
\reference{}
Miller, J. M., et al.~2003, \apj, 583, L99
    
    
\reference{}
Remillard, R. A. 2002, \iaucirc, 7888

\reference{}
Remillard, R. A., Swank, J., \& Strohmayer, T. 2002, \iaucirc, 7893
    
    
\reference{}
Rutledge, R. E., Bildsten, L., Brown, E. F., Pavlov, G. G., \& Zavlin,
V. E., 2001, \apj, 551, 921
         
\reference{}Rutledge, R. E., Bildsten, L., Brown, E. F., Pavlov,
G. G., Zavlin, V. E., Ushomirsky, G., 2002, \apj, 580, 413
    
\reference{}
Rupen, M. P., Dhawan, V., \& Mioduszewski, A. J. 2002, \iaucirc, 7893
    
\reference{}
Smith, D. A., Morgan, E. H., \& Bradt, H. 1997, \apj, 479, L137
               
\reference{}
Stella, L., Campana, S., Colpi, M., Mereghetti, S., \& Tavani,
M. 1994, \apj, 423, L47
    
\reference{}
Stella, L., Campana, S., Mereghetti, S., Ricci, D., \& Israel,
G. L. 2000, \apj, 537, L115
    
\reference{}
Strohmayer, T. \& Bildsten, L. 2004, To appear in 'Compact Stellar
X-ray Sources', eds. W. H. G. Lewin \& M. van der Klis, Cambridge
University Press (astro-ph/0301544)
    
\reference{}
Tomsick, J. A., Gelino, D. M., Halpern, J. P., \& Kaaret, P. 2004,
\apj, 610, 933
    
\reference{}
Wijnands, R. \& van der Klis, M. 1998, \nat, 394, 344
    
\reference{}Wijnands, R., Miller, J. M., Markwardt, C., Lewin,
W. H. G., van der Klis, M. 2001, \apj, 560, L159
                           
\reference{}
Wijnands, R. et al. 2004, \apj, submitted (astro-ph/0310144)       
    
\reference{}
Zhang, W., Jahoda, K., Kelley, R. L., Strohmayer, T. E., Swank, J. H.,
\& Zhang, S. N. 1998, \apj, 495, L9
    
\reference{}Zavlin, V. E., Pavlov, G. G., \& Shibanov, Yu. A., 1996,
\aap, 315, 141
                     	
\end{references}
\end{document}